# Design Challenges of Neural Network Acceleration Using Stochastic Computing

Alireza Khadem
Department of Electrical Engineering and Computer Science
University of Michigan, Ann Arbor, 48109 USA
arkhadem@umich.edu

*Abstract*— The enormous and ever-increasing complexity of state-of-the-art neural networks (NNs) has impeded the deployment of deep learning on resource-limited devices such as the Internet of Things (IoTs). Stochastic computing exploits the inherent amenability to approximation characteristic of NNs to reduce their energy and area footprint, two critical requirements of small embedded devices suitable for the IoTs. This report evaluates and compares two recently proposed stochastic-based NN designs, referred to as BISC (Binary Interfaced Stochastic Computing) by Sim and Lee, 2017, and ESL (Extended Stochastic Logic) by Canals et al., 2016. Using analysis and simulation, we compare three distinct implementations of these designs in terms of performance, power consumption, area, and accuracy. We also discuss the overall challenges faced in adopting stochastic computing for building NNs. We find that BISC outperforms the other architectures when executing the LeNet-5 NN model applied to the MNIST digit recognition dataset. Our analysis and simulation experiments indicate that this architecture is around 50X faster, occupies 5.7X and 2.9X less area, and consumes 7.8X and 1.8X less power than the two ESL architectures.

*Keywords— Neural network acceleration, approximate computing, stochastic computing, design analysis*

## I. INTRODUCTION

Nowadays, neural network is considered as a good way in many hardware use cases such as computer vision and data analytics. As the demand of neural networks increased, many chip designers started to design neural network accelerators among the other ones in their chips. However, with the appearance of convolutional neural networks, the amount of operations and the operators in these models exploded to a high amount. To calculate the result of the models in real-time, designers had to use more and more cores, which conserves a great amount of power as well as the area. In this era, stochastic computing is known as a way of alleviating the mentioned issues by applying the probability laws to digital logic systems. There are many different proposed methods and designs in this regard, which are different in many aspects, such as the way of stochastic number (SN) presentation, processing element (PE) architecture, etc. This research presents a roughly complete survey of the two different SN presentation methods and their corresponding PE architectures, explaining their pros and cons over each other and finally, measures the area, power, and performance of them in practice.

Let us consider the VGG-16 model [1] which consists of 4.3E+12 MAC operation per each input feature map. The MAC logic units in these architectures consume 44% of the on-chip energy. Besides, we require a plethora of MAC units in hardware to implement the neural network models, and then execute them in real-time. To have a comprehension of "real-time" execution, consider processing a 30 FPS video using the VGG-16 model. Thus, we are to process each input feature map in 1/30 second, which requires 1.29E+14 MAC/s. If we implement this network on an FPGA with the clock frequency of 100MHz, neglecting the off-chip access, we need to use 1.3 million MAC units in hardware. Since each MAC unit consists of a multiplier and an adder, which are expensive units in hardware design, this design would be costly in terms of power and area. These are all in the situation that the input feature map of VGG-16 is 224*224 pixels, and today we are processing 4K videos.

With this in mind, stochastic computing is considered as a way to alleviate the aforementioned issues. This computation method is beneficial for the applications that are at first compute-bound and second, amenable to approximation, the two characteristics which are seen mostly at the neural network models.

Stochastic computing encodes data using numerous random variables. Owing to the fact that spatial locality in turbulence in the data affects variables next to each other, using copious random variables reduces the turbulence effect on the number's value.

Furthermore, multiplication is referred to as the most energy and area consuming logic in the neural network acceleration. SC converts this function to simple gates in the stochastic domain. Due to different propagation delay of this logic, performance varies in the accelerators. Besides, diverse formats are of various representation intervals, which also affects accuracy.

Besides, independent numbers in hardware could be manipulated in parallel, which in turn improves the performance. Since an n-bit binary number is converted to many independent bits in the stochastic domain, data could be forwarded in the network separately to the point that it is faced with a nonlinear function. As SC converts complex logic to the simpler ones, and different bits can progress in the network in parallel, hardware architects are able to design the system as a streaming network. Consequently, stochastic designs are of higher performance, occupying the same area.

Nevertheless, the improvements mentioned above are not obtained readily. Although very simple logic serves as a multiplier on the stochastic domain, other essential functions are not the same. For example, addition on this format is counted as a catastrophic drawback. Conventional implementation of this logic Hardware architectures reduces the accuracy, since it scales down the result with a factor of two. One way in this regard is to use other SN presentations, and convert the addition logic to the more accurate ones.

The other problem with the primary stochastic domain is the representation internal, which is [-1 1]. As a result, they are unable to carry the weights which are outside of this interval. Furthermore, limited signal precisioncan deteriorates the accuracy. For example, representation precision of the numbers in the bipolar domain is $\frac{1}{2^N}$, in which N is the number of samples. To improve the precision, we must increase N, which has its adverse effect on the evaluation time (i.e., the amount of time that logic needs in the stochastic network to calculate their final results).

One of the solutions to the mentioned drawbacks is to use other stochastic representations. In the following sections, we probe two different proposed works with three different configurations. The contributions to this paper include:

1. We implement a neural network framework in C language to find the accuracy of different convolutional and fully-connected neural network models. Then, we customize functions in this 1.3K-line framework to be consistent with the chosen stochastic accelerators, and find the accuracy of LeNet-5 network model using three architectures.
2. We implement the hardware description of the three accelerators, simulate them using Modelsim, and synthesize them with Synopsys Design Compiler. Having used different reports obtained from synthesis, we compare the architecture in this paper.

As a result, the following sections are divided as follows: Sec. II gives background information on SC implementation. We will explain our proposed neural network accelerator architecture in Sec. III, while Sec. IV is around the neural network framework and especifically, the neural network model we are implementing. Sec. V and VI discuss the two different works we have chosen to implement on our proposed architecture. In Sec. VII, we are to explain different results we got from implementation and compare the two architectures with each other, and in Sec. VIII we provide a short conclusion.

## II. BACKGROUND

In this section, we introduce different implementations for basic arithmetic operations that we use in the neural network accelerators in the SC format.

### A. Binary and Probabilistic conversion

Data in the network is converted from the binary basis to the stochastic (B2P) represantion to be consistent with the other logic. This conversion mostly occurs using "Stochastic Number Generator" or SNG module. Nevertheless, sometimes designs employ basic logic that converts the basis implicitly. Examples to this logic is APC-based addition [2] and BISC MAC unit [3] which convert data to the binary representation while calculating the output.

An SNG unit consists of a comparator and a Psedo Random Generator function. The input to this module is an N-bit binary number and the output is a $2^N$-bit stochastic number. Each cycle, comparator compares the generated random number with the binary number. The output will be '1' and '0' when the random number is less than and bigger than the binary number, respectively. There are two important aspects when using this function in the accelerator as is mentioned in the next two paragraphs.

Data conversion is one of the error sources in the neural network implementation. This is owing to the fact that the generated random numbers are not completely independent from each other. LFSRs are the pseudo-random generators mostly used in the SNGs. Having considered an N-bit LFSR, this function is able to produce $2^N$ pseudo-random numbers, which will be converted to the $2^N$ bits of the SN in the consequitive cycles. This $2^N$ random numbers will be repeated for the next SNs, which could make correlation for the coming calculations. One way is to using wider LFSR, and shuffling the output sequence for different SNs. The other way is to seeding the LFSRs with different random numbers.

The other important fact is that the comparator compares the N-bit pseudo-random number with the binary number. Thus, designer must convert the binary number to one integer number, to be comparable with the random number. If the output is in the unipolar stochastic format, the input value must range from 0 to $\frac{2^N-1}{2^N}$. However, psedo-randm number generated by the LFSR ranges from 0 to $2^N$-1. As a result, the input value must be multiplied by $2^N$ to be consistent with the random number. If the binary numbers are represented by the fixed-point method with no integer bit and N fraction bits, there is not any extra step to do for this purpose. However, if the output SN is in the bipolar format, binary number must range from $-1$ to $\frac{2^{N-1}-1}{2^{N-1}}$. This is in the situation that the SNG output ranges from 0 to $2^N$. Consiquently, the binary number must be added to 1, and then multiplier by $2^N$.

Data conversion from the probabilistic domain to the binary basis (P2B) is done using down and up-down counters in the unipolar and bipolar formats respectively. The output of the counter must undergo the inverse process descrbed in the previous paragraph.

### B. Multiplication

This operation in the conventional stochastic basis is straightworfard. SC converts this function to simple gates in the stochastic domain, which are *AND*, *XNOR*, and *XOR* logic in unipolar, bipolar, and inverted bipolar formats, respectively.

## C. Addition

Although the data conversion and multiplication are two advantageous of using stochastic computing, addition suffers from the lack of accuracy. Hardware architectures employ multiplexers for addition in the stochastic domain, and the shortcoming of this logic is its scaling factor. Mux-based adders scale down the result with the factor of $f$, which is the number of its inputs. Scaling leads to the information lost. Thus, the designer must use more random samples to preserve the information. Although it seems that this would be a comprehensive solution to this issue, the large number of samples causes a reduction in performance, especially when we have a plethora of input features in the network. The other solution to this problem is using OR logic, which computes $x + y + xy$ when it is fed with two SNs of $X$ and $Y$. However, to reduce the impact of the additional term $xy$, we can train the model in a way that its weights are as small as the result tends to $x + y$.

## D. Activation Function

Nonlinear activation functions are considered as the other disadvantage of the primary representation. Although these formats are able to process the multiplication readily, designing nonlinear activation functions, such as ReLU, is not as easily as multiplication. While neural networks employ both multiplication and nonlinear logic (activation functions), stochastic computing misses its advantages of area and energy to a large extent in this logic. With this in mind, one way is to convert numbers from the stochastic to the binary basis, then calculate the result of the complex functions, and finally bring them back to the stochastic representation. However, conversion between the domains requires extra logic, which has a significant impact on the total energy consumption and performance of the system.

Li *et al.* in [4] are to compare two stochastic neural network accelerators in term of energy-accuracy tradeoff. They have implemented convolutional layers, pooling layers, and fully-connected layers using these two methods of computations.

Unipolar multiplication in both of the architectures is performed by XNOR gates. However, addition is what distinguishes them. The first architecture uses MUX-based addition when the second one employs Accumulative Parallel Counter based or APC-based addition. Although MUX-based addition is efficient in terms of area, it scales the result to $\frac{1}{f}$, in which $f$ is the number of inputs to this adder. APC architecture was proposed [2] to improve the accuracy. This adder computes the total number of ones in the SN, and its output is in the binary representation.

Activation function is the third module described in this paper. It implements hyperbolic tangent using a K-state FSM for SNs. According to [8], this module approximates hyperbolic tangent as $Stanh(K, x) = \tanh(\frac{K.x}{2})$. This type of activation function is used for the MUX-based network. Since the input to this fucntion must be a SN, it is not consistent with the APC-based neuron. Thus, this module employs the scaled hyperbolic tangent activation function. This module is implemented by an up/down counter. Details and information on this function is provided in [5]. Fig X a and b shows the microarchitecture of an APC-based and a MUX-based neurons, respectively.

Although this paper benefits from a thorough comparison in error rate, area, power, and energy, it lacks the evaluation of speedup, which is an essential factor, not only in performance but also in energy consumption.

## III. WEIGHT-STATIONARY CONVOLUTIONAL NEURAL NETWORK ACCELERATOR

There are many approaches to design a processing unit for CNNs. Due to simplicity, weight stationary, input stationary, and output stationary are the famous ones, in which weights, inputs, and outputs of the processing elements are fixed respectively, and the other data change cycle by cycle. Between the three styles, weight stationary seems to be compatible with the architectures in which designers change the weight or the input or output order. I have chosen this approach for all of the three designs in this research. They are all compatible with this method and comparison is much fairer if we use a similar approach for all of these three designs.

Consider the kernel size of *9*. In weight stationary approach, each weight in the kernel is assigned to an exclusive processing element and is multiplied and accumulated by different inputs in different clock cycles. Consider three consecutive clock cycles of this accelerator:

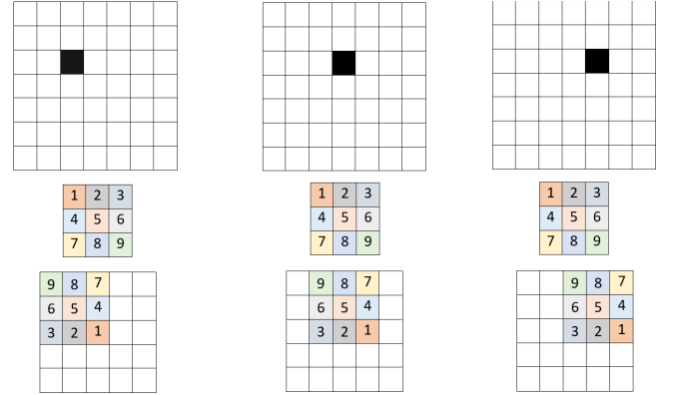

*Figure 1. Three consecutive clock cycles of 21, 22, and 23 in a weight stationary accelerator*

As one can see in Figure 1, the output of *PE#1* is used in the next cycle by *PE#2*, the output of *PE#2* should be routed to *PE#2*, and so on. Thus, the partial sums are carried in one row. The following figure describes the management of the processing elements in the $i_{th}$ row of the kernel. A row of kernel weights are distributed between each processing element, and input is shared between them. Each processing element gets a partial sum from the previous processing element, multiplies its corresponding weight to the input, adds the multiplication result to the previous partial sum, and sends the result to the next processing element. The first and the last processing element fetch/store their partial sum from/to the Partial Result Buffer, which is made from SRAM.

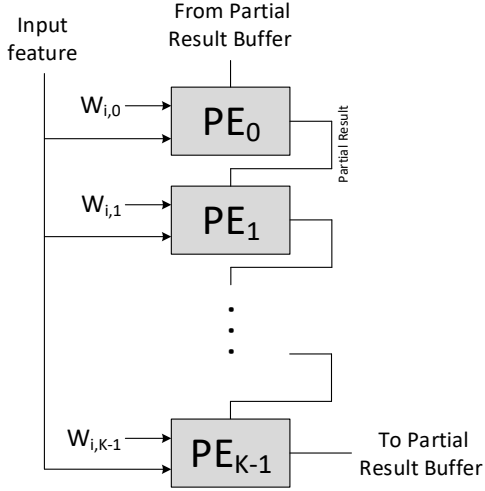

*Figure 2. A row of a weight stationary accelerator*

Now, consider three following inputs in a column. They are processed in three different clock cycles in each row. Figure 3 describes this.

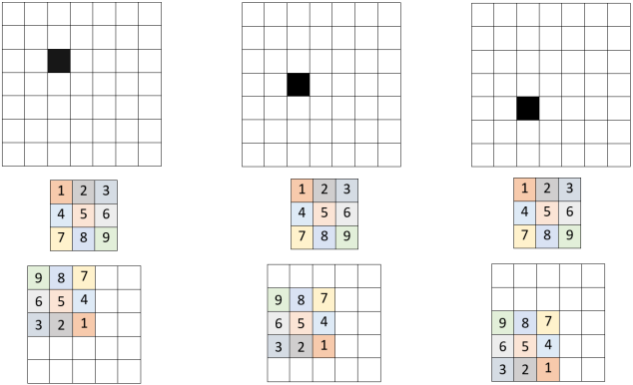

*Figure 3. Three similar clock cycles of 21, 30, and 39 in a weight stationary accelerator*

This figure shows that the output of *PE#3/6* will be used by *PE#6/9* in 9 clock cycles later, which is the width of the input feature map. These partial elements are stored in the Partial Result Buffer to be used in the next row clock cycles. However, there are two exceptions: the output of the processing unit is the output of the *PE#9* (it is not stored, instead, it is routed as the output of the network), and the input to the *PE#1* is 0 (it is not fetched, but it is the constant of 0). The following figure shows the relation between each row and the Partial Result Buffer in a processing unit.

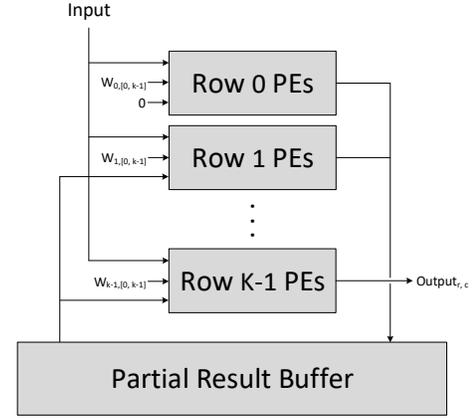

*Figure 4. The relation between the rows of a weight stationary accelerator*

For evaluating the three different architectures which are described in the following sections, we have reused the mentioned accelerator architecture. However, each design has its own customization in the processing unit and the processing element architecutre.

## IV. NEURAL NETWORK FRAMEWORK AND LENET-5 IMPLEMENTATION

We have written a neural network framework to find the accuracy of the neural network models. This framework, which is written in C language, is able to find the accuracy of different convolutional neural networks and fully-connected ones. The input to this program is the weights and the biases of different layers, and input data (images in case of the CNNs, and features in case of FCs.) Since we are using a neural network model which implements an image classification application, the accuracy of the networks are evaluated as the number of correct prediction to the number of test images. The important functions of this framework are as follows:

### A. Data Conversion Functions

For each of the probabilistic to binary and its inverse conversions, we have one function. For binary implementation, these functions just return the input value. For each of the three accelerator we are modeling, we fill these functions with the appropriate functions. (i.e., an up/down counter for P2B, and SNG for B2P conversions)

### B. Basic Functions

We have written one separate function for multiplication, addition, and ReLU activation function. Initially, they are written with the conventional binary operations. We have changed each function with respect to the computation model for the three implementation.

### C. LeNet-5 Function

We have described LeNet-5 model [6] as a top function. This model is a convolutional neural network with 3 CNN lay ers, and 1 FC layer. The inputs are handwritten digit images from the MNIST dataset [7], and the outputs are predictions of the written digits in the corresponding input. Figure 5 shows the architecture of this model.

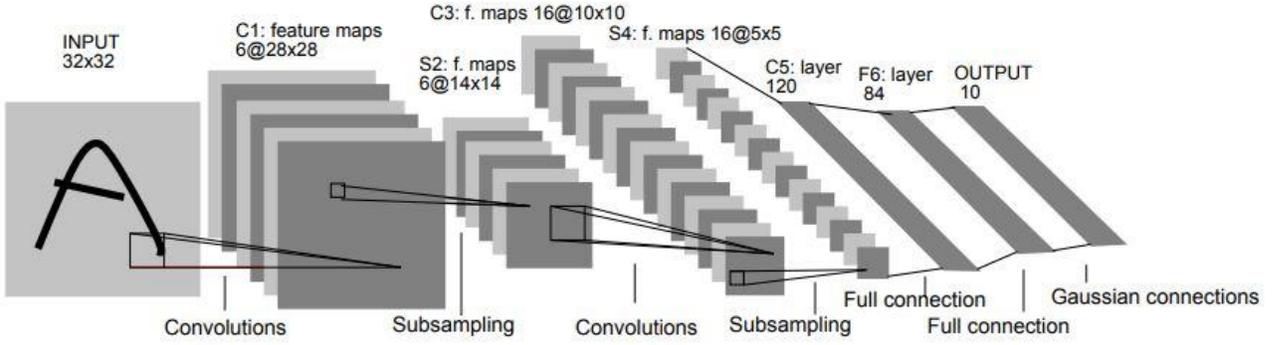

*Figure 5. LeNet-5 neural network model*

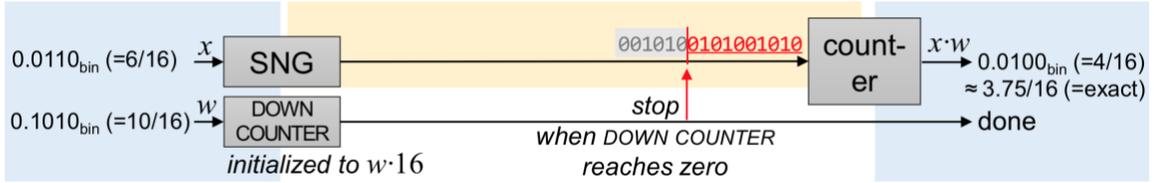

*Figure 6. Low latency multiplication method of BISC*

We extracted the data (weights, biases, input images, and input labels) from a git repository [8] on the internet, and tested our framework with them. The input set includes 1000 test images along with their expected labels from MNIST dataset [7]. We achieved the accuracy of 95.8% using floating point data.

Then, we changed the framework to support the fixed-point data. Figure 7 shows the accuracies achieved with the integer width of 5 and various fraction widths. This figure shows that the accuracy is saturated with the fraction width of 6 bits in this network.

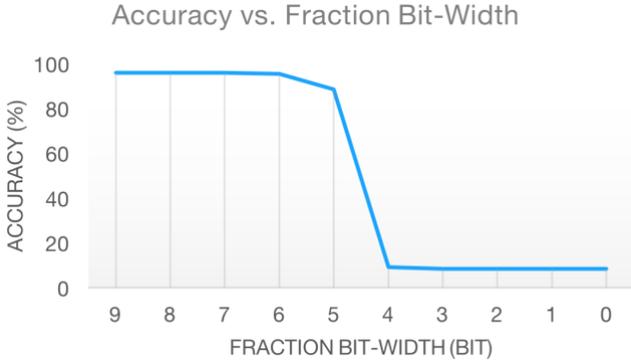

*Figure 7. Accuracy achieved across different fraction bit-widths and the integer bit-width of 5*

In the next step, we found the accuracies of the network using different integer widths. In this experiment, we fixed the fraction bit-width to 6. Figure 8 shows that the accuracy is constant for the integer bit-widths of more than 2 bits. As a result, we chose the integer and fraction widths of 2 and 6, respectively. Thus, this system has 9-bit numbers: 1 bit sign, 2 bits integer, and 6 bits fraction.

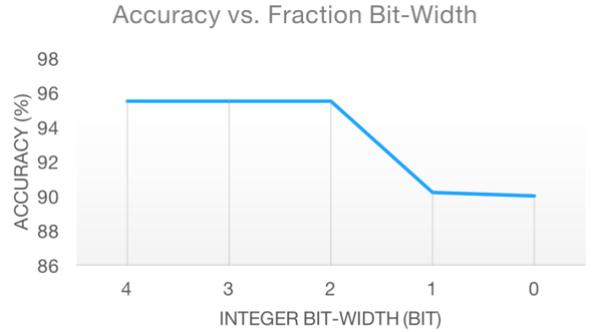

*Figure 8. Accuracy achieved across different integer bit-widths and the fraction bit-width of 6*

V.  FIRST EVALUATED ARCHITECTURE: BISC-MVM
(BINARY INTERFACED MATRIX-VECTOR MULTIPLIER)

Due to long bit-streams, one of the drawbacks of stochastic circuits is their evaluation time. Although the power consumption of these circuits is reduced using simpler logic, long computation time deteriorates energy consumption, and performance as well.

Besides, hardware architects count off-chip memory access as the bottleneck of the neural network accelerators, both in terms of energy consumption and performance, as it consumes around 43% of the total energy in the chip [9]. Moreover, stochastic designs suffer from the high amount of data, since they convert an N-bit wide binary number to an SN with $2_N$ bits to maintain the acceptable accuracy. As a result, data is always stored as a binary number in the off-chip memory, and transformed to the stochastic format using SNGs. To feed the SC neural network accelerator, one should employ a copious number of SNGs, which, in turn, makes the design energy-hungry.

BISC-MVM is proposed by Sim et al. as a way to alleviate the mentioned problem [3]. Consider multiplying $x = \frac{6}{16}$ by $w = \frac{10}{16}$ in the stochastic representation with 16 samples. The stochastic version of w consists of 16 samples with 6 zeros and 10 ones. This approach reorders the samples, such that all the ones come first, and then all of the zeros, which is the unary representation of 10 in 16 digits. BISC-MVM employs AND gate as the multiplier as well as the conventional unipolar stochastic circuits. Owing to the fact that the last 6 digits of *w* are zero, the last six digits of the result will be zero. Thus, the result is equal to the first 10 digits of x, and multiplication can be done using a counter that counts from *w* to zero. By reusing the partial results in the next iterations, one can employ this as a MAC circuit, which the authors call it SC-MAC. Figure 6 shows the aforementioned processing model of MAC operation.

As a result of this kind of SN representation, the multiplication result only depends on the distribution of digits, not the order of them. This fact gives the freedom of choosing the digits even dependent to each other. Having considered $x = \frac{6}{16} = 0.x_3 x_2 x_1 x_0 = 0.0110$ as an example, it is only sufficient to have a string with 8 digits of zero ($x_3$), 4 digits of one ($x_2$), 2 digits of one ($x_1$), and 1 digit of zero ($x_0$). With this in mind, this string could be produced using an FSM and a MUX instead of an SNG, such that the *N-ith* digit (i.e., $x_{N-i}$) first appears at cycle $2_{i-1}$, and then, in every $2_i$ cycles.

As we told in sec. III, we are reusing a unique weight-stationary architecture across the three accelerators with different customizations. In this design, each PE must multiply the inputs by an assigned weight in different cycles, and add the result to the previous partial sums to produce its own partial product. We must consider each PE similar to Figure 4, and customize MAC operation like Figure 6. There are two ways to design this architecture:

*A. First Implementation*

In the first design each PE consists of an up/down counter (counting individual bits of the stochastic number representation of the input number) and a down counter (which counts from w to zero). Consequently, an SNG (producing SN representing input value, or X) can be shared in a processeing unit, and we can broadcast its output across all PEs in the PU. However, number of clock cycles PU needs to calculate the result of the block depends on the value of the different weights assigned to each PE. It means that the evaluation time of each iteration is equal to the maximum value of the weights in the processing unit. This fact increases the number of clock cycles, since the PU must wait for the slowest PE in each iteration before it renews the input.

*B. Second Implementation*

The second implementaion, we have employed, swapps x and w. Thus down counter (shared across all of the PEs) counts from x to 0, and each PE consists of an SNG and an up/down counter. SNG converts the weight to the SN, and the up/down conter evaluates the individual bits in the stochastic representation of w. With this approach, we don't have to wait for the slowest PE, since the evaluation time is equal to the value of x, which is shared across PEs. Figure 9 shows the architecture of a processing element.

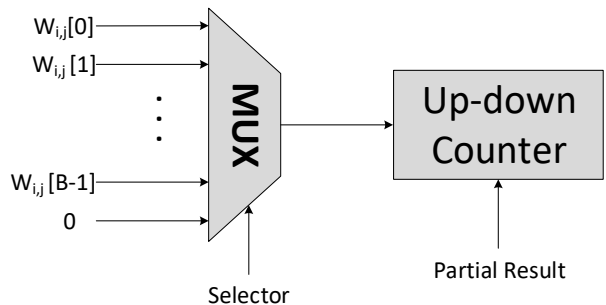

*Figure 9. Processing element in the BISC-MVM architecture*

Each SNG in this architecture contains a MUX and a FSM (which produces the selector of the MUX). FSMs are implemented using a constant memory, in which the bit indices (or selectors) are stored. Since all the PEs are in the same state in one clock cycle, we can share the FSM between the different PEs. Figure 10 depicts the architecture of a PU.

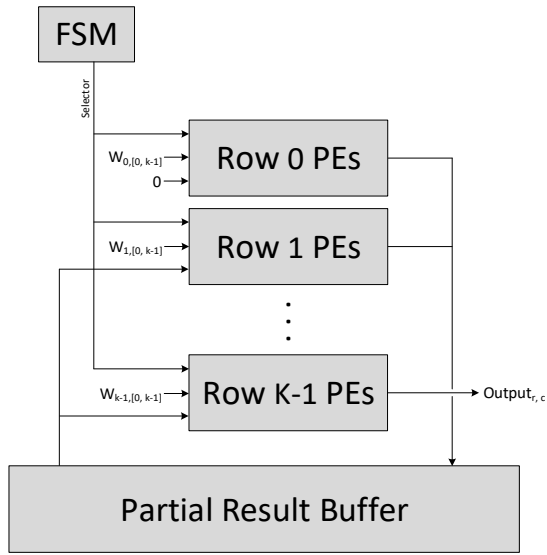

*Figure 10. Processing unit of a BISC-MVM architecture using weight stationary method*

VI. SECOND EVALUATED ARCHITECTURE: ESL (EXTENDED STOCHASTIC LOGIC)

One way to alleviate the addition and interval issues of primary stochastic numbers is to represent a stochastic number with two bipolar stochastic numbers of X and Y with the value of $\frac{x}{y}$.

Multiplication, division, and addition are mapped to their equivalents in the basic stochastic formats, and this is one of the advantages of this method. In the following paragraphs we describe challenges to implement basic opertions in this format.

## A. B2P and P2B Data Conversion

B2P data conversion in this model is simple. As mentioned in Sec. II, for converting bipolar SN to the binary format one can use the SNG module, along with a preprocess on the input number to be compatible with the pseudo-random number generated by the LFSR. However, there is a choice here, and that is how to consider X and Y, such that $\frac{x}{y}$ be equal to the input number. One naïve way is to always consider Y to be 1. The problem with this method is that it is not able to represent the numbers more than 1 or less than -1 since X is a bipolar SN and is in the interval of [-1 1). To solve this problem, for numbers with the value of more than 1, we consider X equal to 1 and Y equal to $\frac{1}{input\ number}$. Although there are some other pairs of X and Y to represent numbers between -1 and 1, this approach is the most accurate one. This is because of the fact that the number of 1s, both in the X and Y, are more than the other approaches, and the SNG has less error when the value of the input number is bigger. This fact will be discussed in Sec. X.

P2B conversion, however, is not as straightforward as be able to use an ordinary up/down counter, since we have two SNs in the ESL. The proposed P2B module in [10] tries to guess the result of $\frac{x}{y}$ in consequetive clock cycles. Thus, we are finding the SN of $P$ such that:

$$P = \frac{X}{Y} \quad (1)$$

As a result, we have:

$$X = Y \times P \quad (2)$$

To guess the amount of P, in the first clock cycle we consider it to be 0. In the consecutive clock cycles, we multiply it by Y, and compare the result with P in the current cycle. If the multiplication result (the right side of (2)) is less than X (the left side of (2)), we increase P by one. However, since more clock cycles are needed for the Ps with bigger values, Akbar et al. have proposed to use the binary search method. This approach we increases/decreases P by $\frac{N}{2^0}$ in the first mismatch, by $\frac{N}{2^1}$ in the second one, etc. One problem with this kind of P2B conversion is that if there is an error in the first bits of the SNs, the output result will be highly erroneous.

## B. Multiplication

This operation in the ESL format is done by two XNOR modules. The following equations shows the multiplication of two ESL numbers.

$$Mul\left(\frac{X}{Y}, \frac{P}{Q}\right) = \frac{X}{Y} \times \frac{P}{Q} = \frac{X \times P}{Y \times Q} = \frac{X \odot P}{Y \odot Q} \quad (3)$$

## C. Addition

One problem with the addition in the conventional MUX-based adders was the scaling factor. This problem is solved in this format. Consider the following addition of two ESL numbers of $\frac{X}{Y}$ and $\frac{P}{Q}$.

$$Add\left(\frac{X}{Y}, \frac{P}{Q}\right) = \frac{X}{Y} + \frac{P}{Q} = \frac{X \times Q + P \times Y}{Y \times Q}$$
$$= \frac{X \odot Q + P \odot Y}{Y \odot Q} \quad (4)$$

According to (4), three XNOR modules and one MUX-based adder are needed to implement the add operation in the ESL format. However, the result of this addition is still suffering from the scale factor issue. Thus, we can convert (4) to two different euqations to eliminate the mentioned problem:

$$\frac{X \times Q + P \times Y}{Y \times Q} = \frac{\frac{1}{2}(X \times Q + P \times Y)}{\frac{1}{2} \times Y \times Q}$$
$$= \frac{MUX(X \odot Q, P \odot Y)}{\frac{1}{2} \odot Y \odot Q} \quad (5)$$

Equation (5) says that by multipying the denominator by the factor of $\frac{1}{2}$, we are able to recover the result, and eliminate the added scale factor. In this implementation, we need a 3-input XNOR module to produce the denominator of the answer. This approach is employed in [10].

In addition, we can use one other MUX-based adder in the denominator and add 0 to it. By this way, both nominator and denominator are scaled down by the factor of 2. This kind of implementation which is explained in (6).

$$\frac{X \times Q + P \times Y}{Y \times Q} = \frac{\frac{1}{2}(X \times Q + P \times Y)}{\frac{1}{2}(Y \times Q + 0)}$$
$$= \frac{MUX(X \odot Q, P \odot Y)}{MUX(Y \odot Q, 0)} \quad (6)$$

Although the described methods try to recover the scale factor, MUX-based addition eliminates half of the information, since it only chooses one sample out of the two samples of its input SNs. Although these approaches rescue the result from the extra factor of half, there is no way to retrieve the eliminated information.

## D. Array Addition

The primary operation used in the neural networks is the MAC operation. A processing element in a binray neural network accelerator multiplies the input by the weight and adds it to the accumulator. If one implements the stochastic neural network in this way (only replace the multiplication and adder with what is described in parts of B and C), the accuracy will have a great error. (This fact is illustrated in the next section.) This is owing to the fact that the two described adders in part B of this section lose the information. Thus, the result of the older multiplications will fade out in the consecutive cycles. We propose three different solutions to this problem, and find the accuracy of them in the next section.

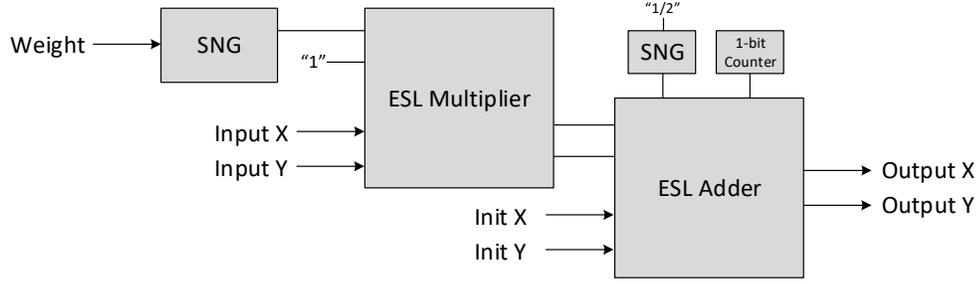

*Figure 11. The processing element architecture of the ESL accelerator*

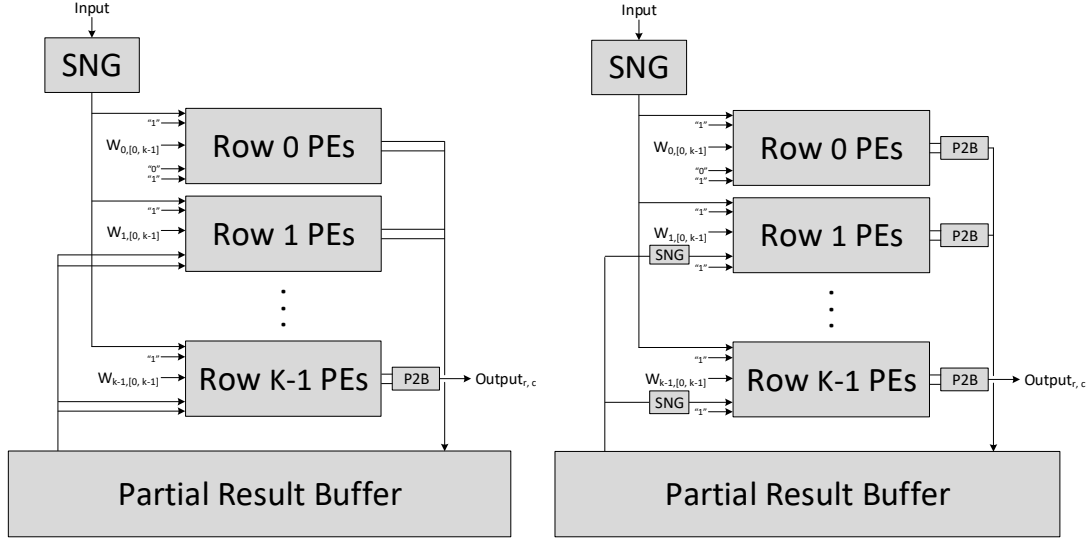

*Figure 12. Two processing unit architectures for the ESL accelerator. The right one stores the partial sums in their stochastic form, the left one converts them to the binary basis and stores/fetches the binary data to/from the Partial Result Buffer*

If instead of adding the numbers sequentially, add them in a binary method such that the operators get at the closest possible distance to the result, information loss will be at its least amount. This is because of the fact that the results of the multiplications undergo the add function less than the sequential approach. This method uses the proposed 2-input adders in the previous part.

The other way is to implement the *f*-input adder, in which *f* is the number of required partial results for an output feature in the neural network. The following equation shows this approach.

$$Add\left(\frac{X_1}{Y_1}, \ldots, \frac{X_f}{Y_f}\right) = \frac{X_1}{Y_1} + \cdots + \frac{X_f}{Y_f} =$$
$$= \frac{(X_1 \times Y_2 \times \ldots \times Y_f) + \cdots}{Y_1 \times \ldots \times Y_f}$$
$$+ \frac{(Y_1 \times \ldots \times X_i \times \ldots \times Y_f) + \cdots}{Y_1 \times \ldots \times Y_f} \quad (7)$$
$$+ \frac{(Y_1 \times \ldots \times Y_{f-1} \times X_f)}{Y_1 \times \ldots \times Y_f}$$

However, designer cannot reuse the same adder in the following clock cycles in the weight stationary approach described in Sec. III. We will evaluate these three array adders in the next section to find their corresponding accuracies.

Figure 11 represents the architecture of a processing element in the ESL design which uses the sequential array adder. As it is displayed in this figure, each processing element includes two SNG units, which is of a great area and power overhead. This overhead shows itself in the area and power reports in the following paragraphs.

Figure 13 is a row of the processing unit. The difference between this design and the basic row processing unit in Figure 2 is that each input/output to/from this accelerator is a pair of bipolar stochastic numbers. (Except for the weight) The other way to design the this architecture is to convert the output to its binary representation using P2B module, and convert it back to the probabilistic basis using B2P modules. This method requires less area and energy to store and load the partial results. Nevertheless, the accuracy falls due to the numerous data conversions.

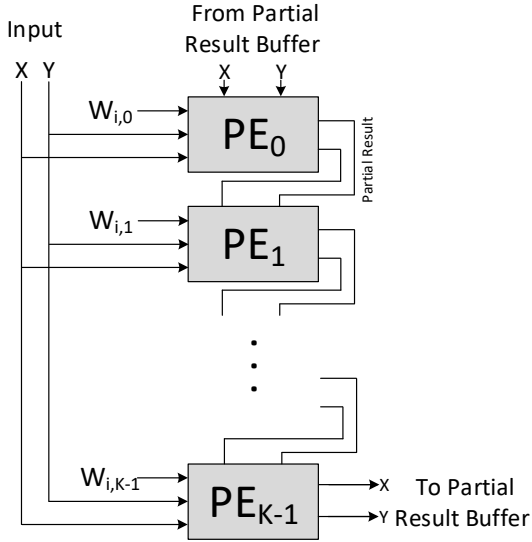

*Figure 13. The architecture of a row of the ESL accelerator*

With all of these in mind, the processing unit could be similar to one of the architectures in Figure 12, depending on the fact that we store partial results in the ESL or binary format. We implement both of them and compare the reports to each other in the next section.

## VII. EXPERIMENTAL RESULTS

We evaluate the three chosen architectures in four parts: performance, power consumption, area, and accuracy. The designs are all described using the Verilog language, and synthesied using a 45nm library. The configuration of the designs are all same to each other, implementing a convolutional layer with the binary width of 6, SC width of 64, kernel width and height of 2, input width and height of 4, and input and output channels of 3 and 4, respectively.

### A. Performance

**BISC-MVM:** The critical path delay is 1.40ns, which means the clock frequency of 714 Mhz. There are 2 critical paths in the system which are listed below.

The first critical path starts from a counter in the FSM, loads the selector from the FSM memory. The selector goes to the PEs and chooses the correct bit of the weight. This bit is accumulated in the up/down counter. The following figure shows this critical path in the system.

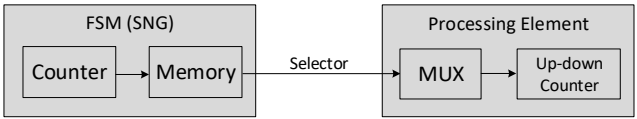

*Figure 14. The first critical path in the BISC-MVM accelerator*

The second critical path starts from the state register in the controller. As the controller is written in the Moore FSM method, the state register is converted to the enable and init signals of the processing units. These two signals go to the Up-down Counters in each processing element, and control the counting in this module. Figure 15 illustrates this critical path.

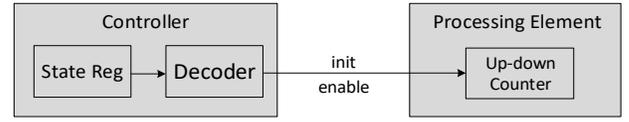

*Figure 15. The second critical path in the BISC-MVM accelerator*

The other factor effective in the performance is the evaluation time for an image to find its category in the LeNet-5 network. Having considered processing this network on just one PE, number of clock cycles in this architecture is equal to the sum of the absolute feature values in the input and partial results. Average sum of the absolute feature values in the 1000 input images is *7.01E+06*. Thus, for each input image this architecture with one processing element needs *7.01E+06cc * 1.40ns = 9.81E-03* seconds to find the output digit on average.

**ESL-without data conversion:** the critical path delay is 2.25 ns, equal to the frequency of 444 MHz. This design has 2 critical paths. The first one starts from the width index counter, goes to the partial result buffer, loads the new init numbers, goes to the processing elements, and adds with the result of multiplication (accumulate operation in MAC units). Figure 16 shows this critical path.

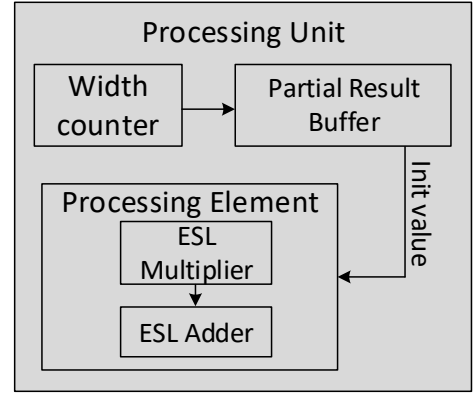

*Figure 16. The first critical path in the ESL accelerator (without partial result basis conversion)*

The other critical path is completely inside processing element module. It starts from the weight SNG, goes to the ESL multiplier, multiplies the weight by the input, goes to the ESL adder, and adds with the initial value which comes from the Partial Result Buffer. Figure 17 depicts this one.

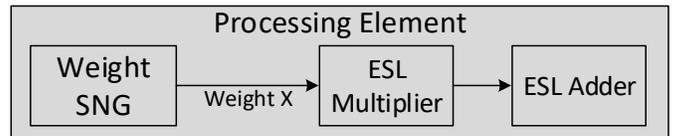

*Figure 17. The second critical path in the ESL accelerator (without partial result basis conversion)*

Since each MAC operation takes 512 cycles in a system with 9-bit wide binary numbers, number of clock cycles needed to evaluate an input image using a processing element is roughly equal to *number of MAC operations × SN length*. Thus, processing LeNet-5 model needs *406800 * 512 = 2.08E+08* clock cycles to produce the output. Multiplying this number by the critical path delay gives us the evaluation time which is *4.68E-01* seconds.

**ESL-with data conversion:** The critical path in this system (the left architecture in Figure 12) is now 2.39 ns, equal to the frequency of 418 MHz. The frequency is slightly less than the previous architecture since now P2B units are in the middle of critical paths, thus, their delay is added to the critical path. The two critical paths are similar to Figure 14 and Figure 15, yet, the P2B modules are added to the end of those paths.

Since this architecture is designed in a way that it hides the time needed for data conversion behind the evaluation time, number of clock cycles needed to evaluate an input image on one processing element is still equal to *2.08E+08*. Multiplying this number by the new critical path delay, gives us the evaluation time of *4.97E-01* seconds.

Figure 18 compares these three architecutres in term of the evaluation time. The values in this figure are normalized to the evaluation time of the ESL with data conversion architecture. In short, BISC-MVM needs 47.6X and 50.6X less time to find the result of an image of the MNIST dataset on the LeNet-5 neural network model compared to the ESL without and with data conversion architectures.

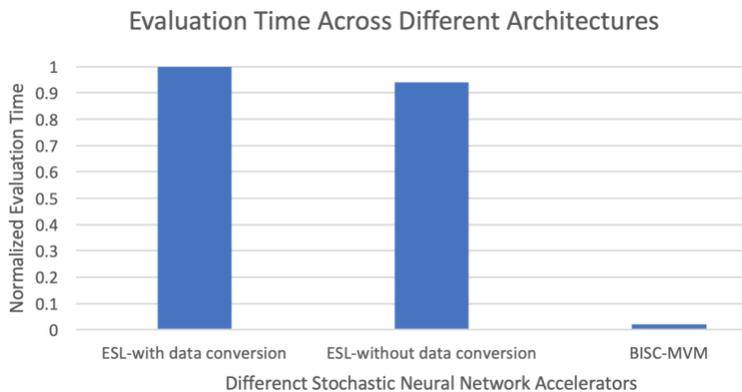

*Figure 18. Evaluation time comparison*

## B. Power Consumption

**BISC-MVM:** The power break-down of this architecture is provided in Table 1.

*Table 1. Power consumption break-down for BISC-MVM architecture*

| Module name | Total Power (mW) | Percentage (%) |
|---|---|---|
| Index counter | 0.309 | 4.1 |
| Partial result buffer | 1.781 | 23.7 |
| Processing elements | 2.979 | 39.5 |
| Selector FSM | 0.702 | 9.3 |
| Down counter | 1.005 | 13.4 |
| Controller | 0.565 | 7.5 |
| Other logic | 0.178 | 2.5 |
| Total | 7.519 | 100 |

As this table shows, around 40% of total power consumption is consumed by the processing elements which consist of the MUXs and up/down counters, and around 24% by the on-chip memory access. Total power consumed by a processing unit with the aforementioned configuration is around 7.5 mW.

**ESL-without data conversion:** Table 2 shows the power consumption breakdown for this architecture.

*Table 2. Power consumption break-down for the first ESL architecture*

| Module name | Total Power (mW) | Percentage (%) |
|---|---|---|
| Index counter | 0.224 | 0.6 |
| Partial result buffer | 28.111 | 77.4 |
| Processing elements | 4.543 | 12.5 |
| SC counter | 0.486 | 1.3 |
| Input SNG | 0.641 | 1.8 |
| Controller | 0.294 | 0.8 |
| P2B convertor | 1.938 | 5.3 |
| Other logic | 7.10E-02 | 0.3 |
| Total | 36.308 | 100 |

Since each data in this format needs 2 * SN length = 1024 bits, its required memory space is 113 times more than the BISC-MVM architecture. As a result, a great part of the total energy consumption (around 78%) is used by the on-chip memory access. Total energy consumption in this architecture for the same processing unit configuration is around 37 mW.

**ESL-with data conversion:** In this design we expect much less power consumption, since data is stored and fetched in the binary format. Table 3, which includes the power consumption break-down for this architecture, shows this fact.

*Table 3. Power consumption break-down for the second ESL architecture*

| Module name | Total Power (mW) | Percentage (%) |
|---|---|---|
| Index counter | 0.212 | 1.5 |
| Partial result buffer | 1.139 | 8.1 |
| Processing elements | 1.173e+01 | 83.8 |
| Input SNG | 0.613 | 4.4 |
| Controller | 0.240 | 1.7 |
| Other logic | 5.50E-02 | 0.5 |
| Total | 1.398e+01 | 100 |

This report shows that partial result buffer power consumption is somehow equal to the BISC-MVM architecture, since both of these designs store data in the binary format. However, processing element power consumption is increased as it now includes the P2B and B2P modules.

As a summary for power consumption, BISC-MVM is less energy hungry than the other architectures. This design consumes 7.82X and 1.85X less power compared to the ESL without and with data conversion architectures, respectively. Figure 19 shows this point. In this figure, power consumptions are normalized to the power consumption of the ESL without data conversion architecutre.

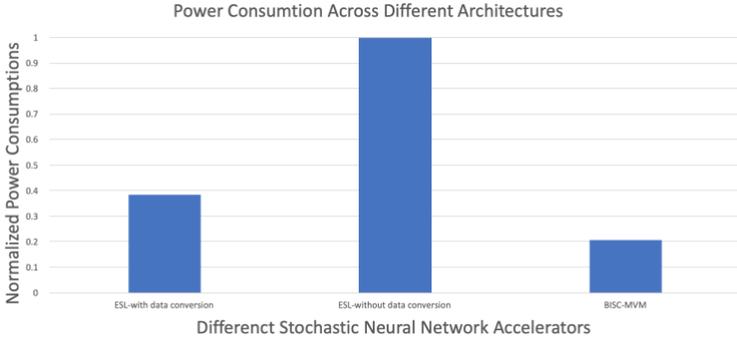

*Figure 21. Power consumption comparison*

## C. Area

**BISC-MVM:** Area break-down of this module is provided in Table 4.

*Table 4. Area break-down for BISC-MVM architecture*

| Module name | Area (μm2) | Percentage (%) |
|---|---|---|
| Index counter | 1.401e-03 | 3.9 |
| Partial result buffer | 8.725e-03 | 24.0 |
| Processing elements | 1.687e-02 | 46.3 |
| Selector FSM | 3.077e-03 | 8.5 |
| Down counter | 2.762e-03 | 7.6 |
| Controller | 2.579e-03 | 7.1 |
| Other logic | 9.652e-04 | 2.6 |
| Total | 3.637e-02 | 100 |

As this table offers, around 47% of total occupied area is used by the processing elements. 24% is occupied by the partial result buffer which is the on-chip memory of the system. Total die area is around 3.7E-02 μm2.

**ESL-without data conversion:** The area break-down of this architecture is summarized in Table 5. As this table offers, a majority of die area is occupied by the partial result buffer (on-chip memory), since it stores data in the ESL format.

*Table 5. Area break-down for the first architecture*

| Module name | Area (μm2) | Percentage (%) |
|---|---|---|
| Index counter | 1.459e-03 | 0.7 |
| Partial result buffer | 1.597e-01 | 76.5 |
| Processing elements | 2.758e-02 | 13.2 |
| SC counter | 2.081e-03 | 1.0 |
| Input SNG | 2.800e-03 | 1.3 |
| Controller | 1.833e-03 | 0.9 |
| P2B convertor | 1.168e-02 | 5.6 |
| Other logic | 1.77E-03 | 0.8 |
| Total | 2.089e-01 | 100 |

Total area needed for this accelerator is around 2.1E-01 μm2.

**ESL-with data conversion:** Since in this architecture we convert data from the ESL format to the binary basis, we need a much less area to accommodate the partial result buffer. Instead, processing elements are bulkier since in this degin, they include the P2B and B2P modules, as well. Table 6 shows the area break-down of the system. Total die area needed for this design is around 1.1E-01 μm2.

*Table 6. Area break-down for the second ESL architecture*

| Module name | Area (μm2) | Percentage (%) |
|---|---|---|
| Index counter | 1.476e-03 | 1.4 |
| Partial result buffer | 9.372e-03 | 8.7 |
| Processing elements | 9.025e-02 | 83.9 |
| Input SNG | 2.579e-03 | 2.4 |
| Controller | 1.708e-03 | 1.6 |
| Other logic | 2.12E-03 | 2 |
| Total | 1.075e-01 | 100 |

In conclusion for this part, BISC-MVM occupies less area since both it stores the data in the binary format (less on-chip memory) and does the data conversion implicitly (less processing element area). Consequently, it occupies 5.7X and 2.9X less area when compared to the ESL without and with data conversion architectures. Figure 19 shows this fact. Numbers are normalized to the area needed for the ESL without data conversion design.

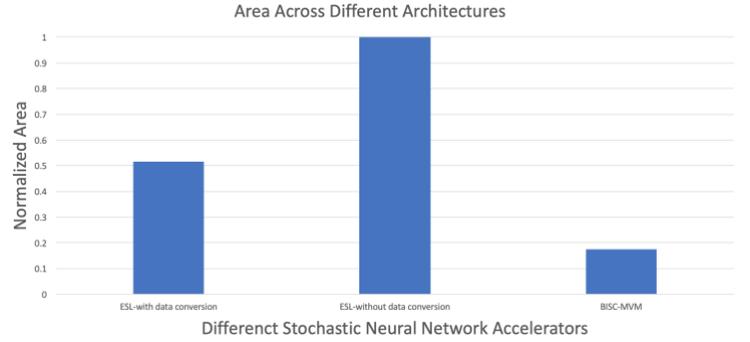

*Figure 19. Area comparison*

## D. Accuracy

**BISC-MVM:** In this design, accuracy is evaluated as the number of correct predictions out of the total number of input images which is 1000 in our framework. Prediction accuracy for this model is 93% while the state-of-the-art accuracy for the fixed-point implementaion is 95.8%. Thus, we have around 3% error while predicting the digits using the LeNet-5 network model.

**ESL- with and without data conversion:** For this design, we evaluate the error of the basic operations. We measure the error using Root Mean Square Error (RMSE) function. Each error is evaluated across 1000 inputs.

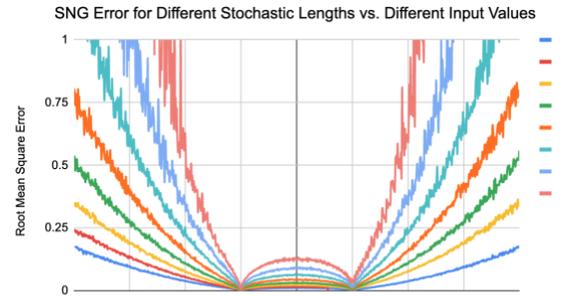

*Figure 20. SNG error across different input values for eight SN lengths*

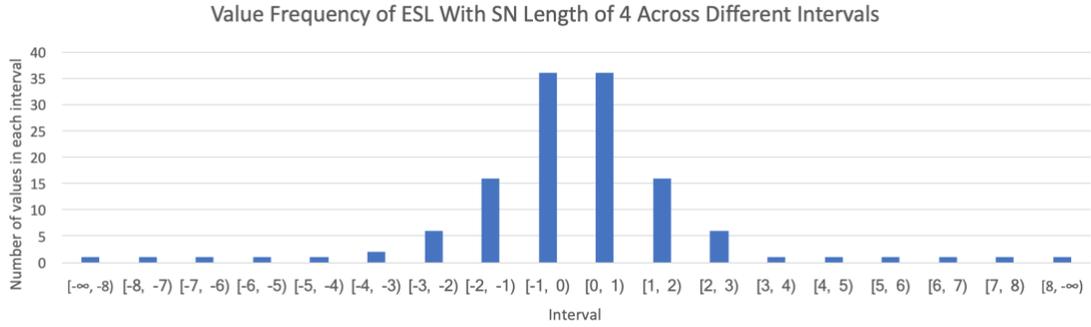

*Figure 22. ESL value distribution*

Figure 21 shows the error of the B2P (SNG) module with different SN lengths ranging from $2_6$ (the smooth dark blue line) to $2_{13}$ samples (the light red line). This figure doesn't show the errors more than 1. As this figure illustrates, when the sample number increases, the error decreases. The RMSE increases at a higher rate for the absolute values bigger than one. Each real value in this representation method is mapped to the nearest ESL candidate in its interval. If we consider the length of the SNs to be 4, Figure 22 represents the frequency of candidates in each interval. As shown in this chart, the frequency of values is not normalized, which in turn can result in different RMSE rate in Figure 21.

P2B error is depicted in figure 23 for different SN lengths. The input value to this module is an ESL number ranging from -1 to 1. Although longer SNs results in lower error in the modules, error is somehow constant across different SN lengths (the orange line with $2_{13}$ samples to the blue line with $2_9$ samples). P2B module tries to guess the value of the ESL number in the consecuitive clock cycles. When the number of samples increase, finding the accurate result gets harder as well. With these in mind, the accuracy remains constant.

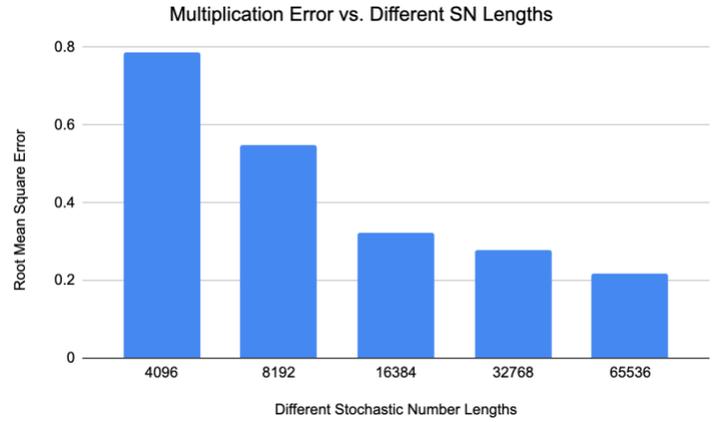

*Figure 23. Multiplication error for different SN lengths*

The RMSE of the multiplier when using 512 samples is on average 6.8. Error across different input values are represented in the 3D chart of Figure 26. The axises of X and Y in this figure are the first and the second input values. The Z axis shows the error of the multiplier. As is obvious in this figure, the multiplier is completely inaccurate in this representation method.

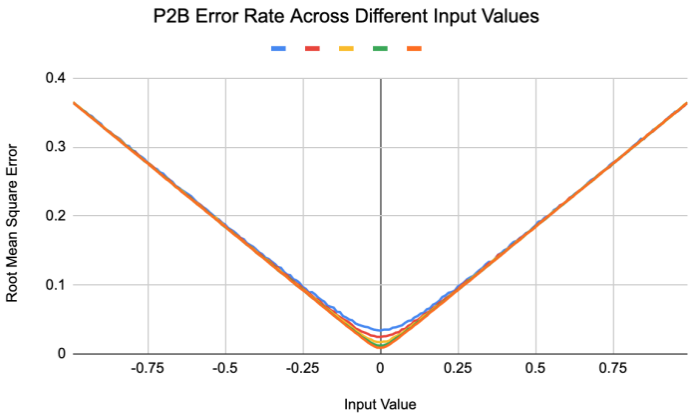

*Figure 24. P2B error for different input values*

Multiplication error is depicted in Figure 22 for various SN lengths, ranging from $2_{12}$ to $2_{16}$. The input values to this function where all in the interval of [-4, 3).

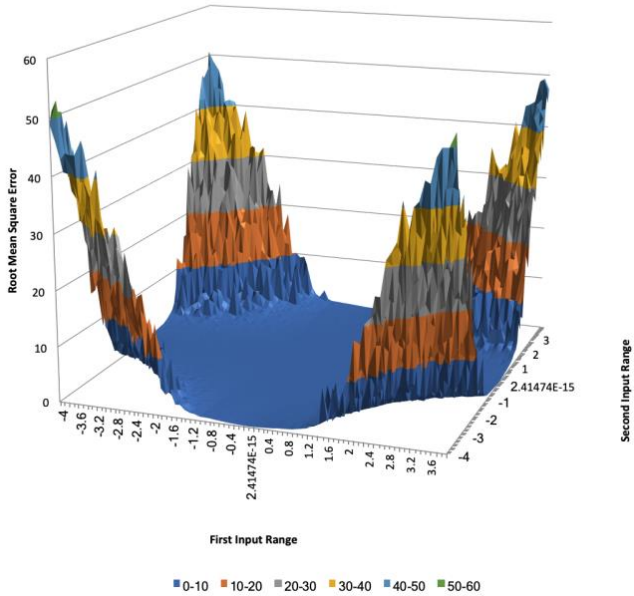

*Figure 26. Multiplication error for different input values*

Figure 29 shows the error of the first array adder architecture described in Ses. VI, part D across different number of inputs to this adder. Number of inputs ranges from 2 to 37, and the RMSE for more than 37 inputs is more than 1 which is not covered by this figure. We have shown the results for different SN lenghts of $2_8$ (the top blue-green line) to $2_{13}$ (the bottom dark blue line). The effect of longer stochastic streams shows itself in this figure as well, when the longer SNs have lower error rate, on average.

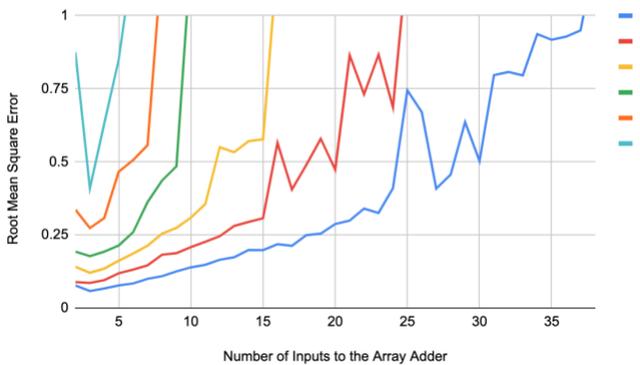

*Figure 29. First adder architecture error for different SN lengths*

Figure 25 shows the same chart for the second Array Architecture, while Figure 27 shows the result for the third architecture. Number of inputs to these adders ranges from 2 to 6, since feeding these adders with more than 6 operands the will produce an error higher than one. Besides, the stochastic lengths in these charts starts from $2_{10}$ (the top green line) to $2_{13}$ (the bottom blue line). If we feed these modules with the SN length of less than $2_{10}$ bits, the error will be more than 1, which is not depicted in the charts. Having compared these charts with the one in Figure 29, one can easily say that the accuracy of the first implementation is more than the second and the third implementations. Thus, we have employed the first architecture both in the C framework and the HDL code.

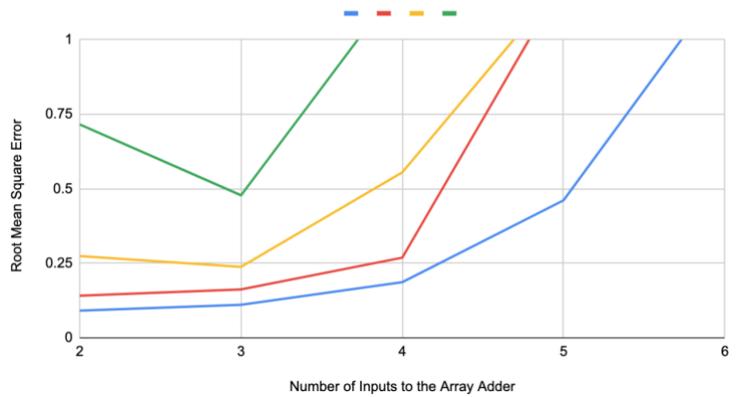

*Figure 25. Second adder architecture error for different SN lengths*

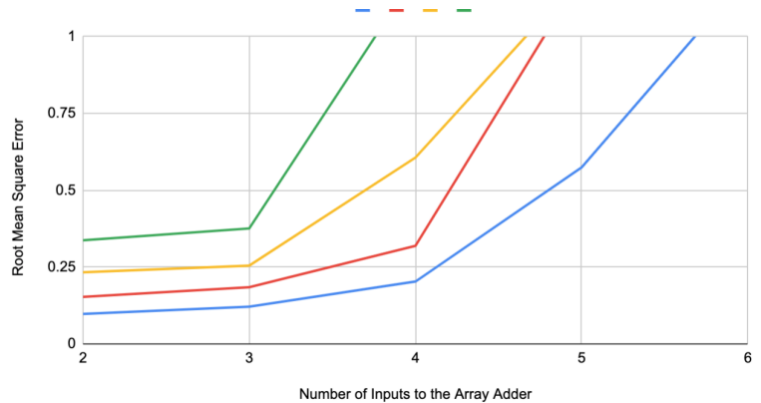

*Figure 27. Third adder architecture error for different SN lengths*

With respect to the high errors in the previous figures, when using $2_9$ bits, the accuracy of the ESL architecture without data conversion is 19.3%, while in the architecture with data conversion the accuracy falls to the value of 13.2%. Figure 28 compares the accuracies of these three architectures running the LeNet-5 neural network model. Numbers in this figure are normalized to the accuracy of the BISC-MVM architecture.

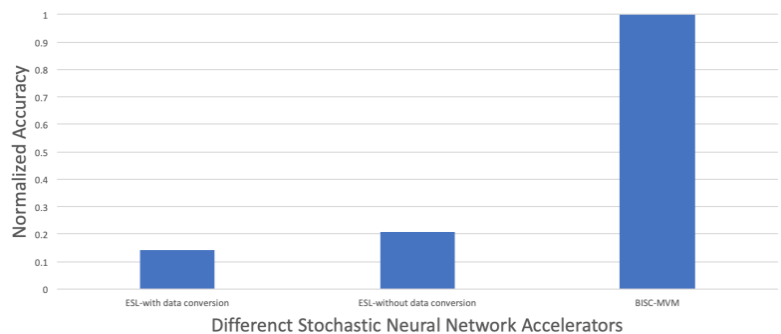

*Figure 28. Accuracy comparison*

## VIII. Conclusion

In this study, we compared two stochastic-based NN designs using three different architectures in terms of performance, power consumption, area, and accuracy. For the second accelerator, which is ESL, we considered two different architectures. The first one stores partial results to memory in the binary format, while the second one stores them in their raw ESL format. Our results showed that the design with data conversion is 6% slower and 6.1% less accurate, while occupies 1.96X less area and consumes 4.22X less power.

We proposed a novel weight-stationary neural network accelerator and implemented the three chosen architectures based on this work. The accelerator is written in a way that a designer can change individual modules in the HDL code to make it consistent with the other methods of computation. For a fair comparison, the implementations of three architectures are completely similar to each other, with some minor modifications in the PE and PU modules. We simulated and synthesized the architectures using a 45nm cell library, and according to the synthesis reports, BISC is around 50X faster than the two ESL architectures. It also comes with 5.7X and 2.9X less area, and 7.8X and 1.8X less power.

Finally, we wrote a C framework to find the accuracy of a CNN, and applied the three architectures to the code to find their accuracy when guessing the MNIST digits. Our results showed that BISC with 93% accuracy is the most accurate architecture, as well. We have evaluated the accuracy of different basic operations in the ESL format and also compared three different array adders in this representation method in terms of accuracy. Our reports illustrated that the parallel array of adders is more accurate than the other architectures.

Consequently, we conclude that between the three SC-based neural network architectures that we evaluated, BISC-MVM outperforms the ESL implementations when executing the LeNet-5 NN model in terms of accuracy.